\documentstyle[12pt] {article}
\newcommand{\beq}{\begin{equation}}
\newcommand{\eeq}{\end{equation}}
\newcommand{\beqa}{\begin{eqnarray}}
\newcommand{\eeqa}{\end{eqnarray}}
\newcommand{\ba}{\begin{array}}
\newcommand{\ea}{\end{array}}

\begin{document}

\def \mtil{{\tilde m}}

\begin{center}
{\bf SPECTRAL DISTRIBUTION STUDIES WITH MODIFIED KUO-BROWN INTERACTION \\
IN THE UPPER HALF OF THE fp SHELL} \\
\vskip 0.3 truecm
{\bf S. Choubey} and {\bf K.Kar}\\
Saha Institute of Nuclear Physics \\
Block - AF, Sector I, Bidhan Nagar \\
Calcutta 700 064, India \\
\vskip 0.2 truecm
{\bf J.M.G. Gomez} \\
Departamento de Fisica Atomica, Molecular y Nuclear \\
Facultad de Ciencias Fisicas, Universidad Complutense de Madrid \\
E-28040 Madrid, Spain \\
\vskip 0.2 truecm
{\bf V.R. Manfredi} \\
Dipartimento di Fisica "G. Galilei", Universit\`a di Padova \\
Istituto Nazionale di Fisica Nucleare - Sezione di Padova \\
via Marzolo 8, I-35131, Padova, Italy \\
\vskip 0.25 truecm
\centerline{Received: \underbar {~~~~~~~~~~~~~~~~~~~~~~~~~~~~~~~}}
\end{center}
\vskip 0.1 truecm

\begin{center}
{\bf ABSTRACT}
\end{center}

\vskip 0.5 truecm

{\it The spectral distribution method with modified Kuo-Brown interaction 
is extended to the study of the upper half as well as to 
odd-A nuclei, of the fp shell. The calculations show similar success to that 
obtained for the lower half.} 
 
\newpage

Microscopic calculations in the fp-shell involving all four valence 
orbits $f_{7/2}$, $f_{5/2}$, $p_{3/2}$ and $p_{1/2}$ have been 
actively persued in the last few years. Among the residual two body interactions 
used for this purpose modified Kuo-Brown (KB3) has shown remarkable 
success for the lighter fp-shell nuclei 
in the full shell model diagonalisation calculations [1,2] 
as well as in the Monte-Carlo shell model studies [3] for both spectra 
as well as transition strengths. Spectral distribution theory, originally 
constructed to reproduce the global features of level densities and 
transition strengths [4] is seen also to reproduce binding energies, 
low lying spectra and transition strengths equally well. These  
studies were performed in detail for the sd-shell 
[4,5] and in some cases for the fp-shell nuclei [6]. Recently such 
calculations were carried out with the KB3 interaction in the lower 
half of the fp-shell [7] with reasonable success. Some of the 
studies are motivated by applications to nuclear astrophysics, such as  
the problem of calculating the electron capture/beta decay rates 
for supernova and presupernova evolution and r and s process 
nucleosynthesis. Therefore one feels the need to extend the calculations to 
the upper half of the fp-shell as many important nuclei for such 
applications have A$>$60. We should note here that KB3 was constructed to 
improve the spectroscopy of nuclei at the beginning of the fp-shell and in 
the absence of full diagonalisation the success of KB3 for heavier fp-shell
nuclei is still uncertain. But KB3 correctly modifies the diagonal matrix 
elements of the 2-body residual interaction and as spectral distribution 
theory relies on the lower order moments of the Hamiltonian in the many body 
shell model spaces, which depend more crucially on the diagonal matrix elements 
than on the nondiagonal ones, we consider it interesting to apply the KB3 
interaction for the spectral distribution studies in the upper half of 
the fp-shell. In this report we study  
spectra, 
occupancies and sum rule strengths of transition operators, using the 
KB3 interaction in the upper half of the fp-shell. Here we 
incorporate the corrections coming from the third and the fourth 
moments of the Hamiltonian in terms of the nonzero skewness ($\gamma_1$) 
and excess ($\gamma_2$). We shall see that compared to the earlier spectral 
distribution results [6], the agreement of 
the binding energies with the experimental values, particularly for 
nuclei with A$>$70, is considerably improved. We also apply KB3 interaction to the odd-A 
nuclei in the fp-shell for the first time and see that after taking 
into account corrections from nonzero ($\gamma_1$,$\gamma_2$) the 
agreement with experimental data is satisfactory.

Spectral distribution theory gives smoothed fluctuation free forms for the 
density of states as a function of energy, which in large shell model 
spaces asymptotically go towards gaussians. The formal derivation 
of the result uses the central limit theorem (CLT) for the one body 
Hamiltonian (H(1)) and extends that to the two body Hamiltonian (H(2)), 
defining an ensemble of Hamiltonians and averaging the moments $<H^p(2)>$, 
p=1,2,...over the ensemble [8]. For the 2-body Hamiltonian the ensemble 
averaged results in many particle space follow from the dominance of binary 
correlations, as elegantly demonstrated by Mon and French [8].  
Spectral distribution also gives 
polynomial expansions for the expectation values of operators in terms 
of energy where only the first two terms in the expansion contribute 
in the CLT limit [4]. Partitioning the shell model space of m valence 
particles in N single particle states according to configuration and 
isospin ($(\mtil,T)$ spaces where $\mtil = m_1, m_2,...m_l$ are the 
number of particles in the l orbits) and the use of gaussian or 
gaussian modified by Cornish-Fisher expansion around it for the density of states, increases 
the predictability of the method. To find the ground state energy 
$\bar{E_g}$ one inverts the equation
$$
\sum_\mtil \int_{-\infty}^{\bar{E_g}} I_{\mtil,T}(E)dE=d_0/2
\eqno (1)
$$
where $I_{\mtil,T}$ is the gaussian density of states in the ($\mtil,T$) 
spaces normalised to d($\mtil,T$), the dimensionality of the configuration 
isospin space, and $d_0$ (=(2J+1)) is the degeneracy of the ground state 
with spin J. Thus one integrates the area below the gaussian configuration 
densities until the area becomes equal to half the ground state degeneracy. 
This energy value is the predicted ground state energy and the method is 
called the Ratcliff procedure [9]. 
As done earlier for the sd-shell [10] and the lower half of the 
fp-shell [7], we improve the predictions by incorporating the corrections 
from nonzero skewness ($\gamma_1$) and excess ($\gamma_2$) by using the 
Cornish-Fisher expansion which gives
$$
x=y+{{\gamma_1}\over{6}}(y^2-1)+{{\gamma_2}\over{24}}(y^3-3y)-{{{\gamma_1}^2}\over{36}}(2y^3-5y)
\eqno (2)
$$
where $y$ is the normalised energy ($y=(E-\epsilon)/\sigma$, $\epsilon$ is the 
centroid and $\sigma$ is the width) before the 
correction and $x$ is the value after it. For the centroids and widths we use 
the values in (m,T) spaces. For $\gamma_1$ and $\gamma_2$, ideally one 
should use the values in (m,T) spaces calculating the third and the fourth 
moments of the (1+2)-body Hamiltonian. But as at present the 
spectral distribution method (SDM) codes 
can calculate them only in the scalar (m) spaces. We use a phenomenological 
correction term for the excess using $\gamma_2=\gamma_2(m)+am+bm^2$ 
where the values of 'a ' and 'b' are obtained through a best fit. In the upper 
half of the fp shell, we find that a parametrised dependence of $\gamma_2$
on isospin makes little improvement in the agreement of the calculated
binding energies with the experimental ones in contrast to the lower half. As the
$\gamma_1$ corrections are very small, we keep the $\gamma_1(m)$ unchanged. 
We calculate the binding energies at fixed configurations before and after 
the corrections and give in Table 1 the difference between the predicted and the 
experimental values given by DIFF=(calculated binding energy - experimental 
binding energy) MeV. The binding energies are taken to be positive in 
agreement with the convention used by experimentalists. We also list the 
corresponding values obtained by Haq and Parikh [6] to show that by using the KB3 
interaction and by incorporating the  
($\gamma_1,\gamma_2$) corrections, we obtain considerable improvement 
over the earlier calculations done by them, where they utilise scalar isospin 
moments with excited state correction and use the MHW2 interaction. 
For the odd-A nuclei the best predictions are obtained using a=0.006 and 
b=-0.00029 and the RMS deviation of the calculated values from the 
experimental values is 1.22 MeV.
For the odd-odd nuclei the best predictions are obtained using a=0.005 and 
b=-0.00028 and the RMS deviation of the calculated values from the 
experimental values is 1.34 MeV. 
For the even-even nuclei the best predictions are obtained using a=0.007 and 
b=-0.00031 and the RMS deviation of the calculated values from the 
experimental values is 1.77 MeV. One observes that correction of $\gamma_2$
separately for the odd-A, even-even and odd-odd nuclei brings the values
closer to experimental ones, as this way one is able to take account of the ground
state pairing effects to some extent.  

There are indications from experimental pick-up/stripping reaction data that the orbit 
$1 g_{9/2}$ starts picking up neutrons when the neutron number of the nuclei 
goes close to 40 [11]. We have constrained our calculations to the four fp-shell 
orbits and this may be one of the reasons for the somewhat larger deviations 
of the SDM values from the experimental binding energies for neutrons almost 
filling the shell.  
We note the considerable improvement one achieves by 
indroducing the ($\gamma_1, \gamma_2$) corrections for these nuclei compared 
to the Haq-Parikh values as seen in Table 1. 
But we feel future studies for A=70 and beyond 
should include excitations of particles to the $1 g_{9/2}$ orbit.

To understand the global properties of the KB3 interaction one should study 
the centroids, widths and its correlation coefficient with other typical 
interactions in the fp-shell evaluated in the (m,T)  
spaces. For the sake of comparison, we choose the MHW2 interaction [12]. We give 
the typical example of m=24 and m=28 with their two extreme isospin values,but 
the behavior at other particle numbers and isospins is very similar. For 
m=24, the centroids for T=0 and T=8 are -243.89 MeV and -193.21 MeV  
for KB3 and -229.80 MeV and -186.49 MeV for MHW2 respectively. For m=28 the 
centroids for T=0 and 6 are -304.12 MeV and -274.55 MeV for KB3 and -285.25 MeV 
and -259.98 MeV for MHW2 respectively.As essentially KB3 differs from MHW2 in 
the diagonal two-body matrix elements, one understands the differences in the 
centroids observed. The widths for the MHW2 interaction are seen to be 
different from the KB3 interaction by a factor between 0.75 to 0.90 in the 
upper half. For example for m=24 and T=0 the width for KB3 is 14.89 MeV, 
whereas for MHW2 it is 13.09 MeV. As the nondiagonal elements for the two 
interactions are almost identical, the correlation coefficient between the 
two interactions, which has the centroids subtracted out, is always close to one 
throughout the shell. Typically for m=28 it varies between 0.992 to 0.989.

Spectral distribution gives a polynomial expansion in energy for the expectation 
values of operators where the terms beyond the first two in the expansion 
are inhibited by CLT. Explicitly for the operator $K$ in the (m,T) space, the 
CLT form for its expectation value at energy E is,
$$
K(E;m,T)=\langle m,T|K|m,T\rangle + \zeta_{K-H}(m,T)\sigma_K(m,T){{E-E_c(m,T)}\over{\sigma(m,T)}}
\eqno (3)
$$
where $\sigma_K(m,T)$ is the width of the operator $K$ in the space (m,T)  and 
$\zeta_{H-K}$ is the correlation coefficient between $K$ and $H$ in the space (m,T). 
$E_c$(m,T) and $\sigma(m,T)$ are the centroid and width of the Hamiltonian in 
the same space. Taking $K=n_s$, the number operator for the orbit s, one can find 
the occupancy of the orbit s. One can also obtain such expansions for the 
expectation values in specific configuration isospin ($\mtil$,T) spaces and then 
average over all configurations. Table 2 shows such configuration 
averaged occupancies calculated for the 4 orbits at the ground state energies 
for some typical values of valence nucleons and isospins. 
These results are also available for all valence nucleon numbers and isospins. 
But we mention here that the SDM values given are averaged over all J states. 
Occupancies through shell model calculations or from experimental 
pick-up/stripping sum rules are for spaces with fixed J. Ideally one 
needs to do a J-projection of the SDM value for proper comparisons.  
A detailed comparison of such occupancies by SDM and the shell model in 
the fp-shell, is, in our opinion, of great interest.

 We also use for $K$ the Gamow-Teller (GT) sum rule operator i.e. $K=(O^{11}\times O^{11})^{00}$ 
with $O^{11}$ being the GT transition operator, a vector in both J and T. 
Then for states with isospin zero, we obtain through Eq.(3) the sum rule 
strength for the Gamow-Teller transition to all final states. In Table 3 we 
give sum rule strengths for self conjugate nuclei (i.e. with N=Z ) with 
valence particles 22, 24, 32 and 34. The table explicitly shows how the correlation 
of $K$ with $H$ changes the sum rule strengths. As the correlation coefficient between 
$K$ and $H$ is very small, the inclusion of the second term brings about a decrease 
of less than 3$\%$. These sum rule estimates of the GT strength are useful for the 
calculations of the electron capture rates on these nuclei during the collapse 
phase of the supernova or the beta decay rates for the presupernova evolution [13].

One can also use such sum rule estimates for other one body interaction operators 
also. These calculations can be extented to nuclei with nonzero ground state 
isospin. The success of the SDM in reproducing average energies as well as 
transition strengths gives one a method of evaluating many structural properties 
of nuclei with many valence nucleons in active orbits, avoids explicit 
diagonalisation and is also useful for astrophysical applications.

We thank S. Sarkar for his help and for many useful discussions; 
V. K. B. Kota for supplying us some of the SDM  programs.
This work has been partially supported by the Secific Project "Milano 41" 
and DGES Project No. PB96-0604.

\newpage
{\bf REFERENCES}

\vskip 0.5 truecm

[1] E. Caurier, A.P. Zuker, A. Poves and G. Martinez-Pinedo,
    Phys. Rev. C 50, 225 (1994); E. Caurier, G. Martinez-Pinedo, 
    A. Poves and A.P. Zuker, Phys. Rev. C 52, R1736 (1995).

[2] G. Martinez-Pinedo, A. P. Zuker, A. Poves and E. Caurier, 
    Phys. Rev. C 55, 187 (1997).
   
[3] S.E. Koonin, D.J. Dean and K. Langanke, Phys. Rep. 278, 1 (1996); 
    K. Langanke, D. J. Dean, P. B. Radha, Y. Allhassid and S. E. Koonin, 
    Phys. Rev. C 52, 718 (1995).

[4] J.B. French in "Nuclear Structure", 1967, edited by A. Hossain, 
    Harun-al-Rashid and M. Islam, (North-Holland, Amsterdam); 
    J.P. Draayer, J.B. French and S.S.M. Wong, Ann. Phys. (N.Y.) 
    106, 472 (1977); J.B. French and V.K.B. Kota Annu. Rev. Nucl. 
    Part. Sci. 32, 35 (1982).

[5] V.K.B. Kota and K. Kar, University of Rochester, Laboratory 
    Report No. UR-1058, 1988; Pramana 32, 647 (1989).

[6] R.U. Haq and J.C. Parikh, Nucl. Phys. A273, 410 (1976); 
    V.K.B. Kota and V. Potbhare, Nucl. Phys. A331, 93 (1979).

[7] K. Kar, S. Sarkar, J.M.G Gomez, V.R. Manfredi and L. Salanich, 
    Phys. Rev. C 55, 1260 (1997).

[8] K. K. Mon and J.B. French, Ann. Phys. 95, 90 (1975).

[9] K. F. Ratcliff, Phys. Rev. C 3, 117 (1971).

[10] S. Sarkar, K. Kar and V.K.B. Kota, Phys. Rev. C 36, 2700 
    (1987).

[11] G. J. Matthews, S. D. Bloom, G. M. Fuller and J. N. Bahcall, 
    Phys. Rev. C 32, 796 (1985).

[12] J. B. McGrory, B. H. Wildenthal and E. C. Halbert, Phys. Rev 
     C 2, 186 (1970).

[13] K. Kar, A. Ray and S. Sarkar, Astrophys. J. 434, 662 (1994)

%ooooooooooooooooooooooooooooooooooooooooooooooooooooooooooooooo
\newpage
\begin{center}
{\bf Table 1}
\end{center}

The difference, DIFF in MeV, between the experimental and the calculated 
binding energies. 
Column $\bar A$,$\bar B$ and $\bar C$  
give the difference corresponding to the value calculated 
by Ratcliff procedure in $(\mtil,T)$ space, 
by Ratcliff procedure  
with ($\gamma_1(m)$,$\gamma_2(m)$) correction and 
by Ratcliff with ($\gamma_1(m,T)$,
$\gamma_2(m,T)$) correction respectively. The last column gives DIFF corresponding to 
earlier SDM predictions with excited state correction using the MHW2 
interaction [6].

$$\halign{&\hfil\  # \ \hfil \ \cr
\noalign{\hrule}\cr
\hfil Nucleus 
&\multispan{3}\hfil DIFF \hfil&DIFF\cr
    & \multispan {3} \hfil with KB3 \hfil & with  \cr
  & $\bar A  $ &$ \bar B$&$ \bar C$& MHW2 \cr
\noalign{\hrule}\cr
$^{61}Zn$&18.3&-0.6&-0.8&4.0\cr
$^{61}Cu$&18.1&-0.5&-0.7&2.0\cr
$^{61}Ni$&17.6& 0.2& 0.1& 0.9\cr
$^{61}Co$&15.1& 1.1& 0.9& 1.7\cr
$^{61}Fe$&14.4& 2.5& 2.4& 3.1\cr
$^{63}Ga$&19.2& 1.1&-0.6& 2.5\cr
$^{63}Zn$&19.5& 1.5&-0.1& 3.0\cr
$^{63}Cu$&18.0& 1.2&-0.4& 0.5\cr
$^{63}Ni$&17.7& 2.1& 0.7& 1.8\cr
$^{63}Co$&16.6& 5.1& 4.0& 1.4\cr
$^{65}Ge$&19.9& 3.1& 0.1& 0.6\cr
\cr
\noalign{\hrule}\cr}$$

\newpage

{\bf Table 1 (contd.)}
$$\halign{&\hfil\  # \ \hfil \ \cr
\noalign{\hrule}\cr
\hfil Nucleus 
&\multispan{3}\hfil DIFF \hfil&DIFF\cr
    & \multispan {3} \hfil with KB3 \hfil & with  \cr
  & $\bar A  $ &$ \bar B$&$ \bar C$& MHW2 \cr
\noalign{\hrule}\cr
$^{65}Ga$&19.6& 2.9&-0.1& 0.4\cr
$^{65}Zn$&17.8& 2.6&-0.1&-3.9\cr
$^{65}Cu$&15.9& 2.1&-0.4&-5.2\cr
$^{65}Ni$&12.6& 1.8&-0.2&-8.4\cr
$^{67}Ga$&17.7& 3.9& 0.0&-4.6\cr
$^{67}Zn$&14.6& 3.0&-0.3&-10.2\cr
$^{67}Cu$&11.9& 2.1&-0.7&-10.9\cr
$^{69}Ge$&15.9& 4.8& 0.5&-9.6\cr
$^{69}Ga$&13.7& 3.7&-0.1&-13.8\cr
$^{69}Zn$&11.6& 3.2&-0.1&-15.0\cr
$^{69}Cu$& 4.6&-0.4&-2.3& 22.2\cr
$^{71}Se$&15.7& 6.3& 1.8&-12.4\cr
$^{71}As$&13.9& 5.4& 1.3&-16.6\cr
\cr
$^{62}Zn$&20.2&-0.4& 0.1& 2.1\cr
$^{62}Ni$&18.2& 0.2& 0.7& 0.1\cr
$^{64}Ge$&19.9& 1.1& 0.0& 2.0\cr
$^{64}Zn$&20.1& 1.3& 0.2&-0.3\cr
$^{64}Ni$&15.9& 1.2& 0.3&-3.7\cr
\cr
\noalign{\hrule}\cr}$$

\newpage

{\bf Table 1 (contd.)}
$$\halign{&\hfil\  # \ \hfil \ \cr
\noalign{\hrule}\cr
\hfil Nucleus 
&\multispan{3}\hfil DIFF \hfil&DIFF\cr
    & \multispan {3} \hfil with KB3 \hfil & with  \cr
  & $\bar A  $ &$ \bar B$&$ \bar C$& MHW2 \cr
\noalign{\hrule}\cr
$^{66}Ge$&21.2& 3.3& 0.5& 0.7\cr
$^{66}Zn$&18.2& 2.5& 0.1&-5.2\cr
$^{66}Ni$&12.1& 1.4&-0.3&-8.7\cr
$^{68}Ge$&19.3& 4.3& 0.4&-4.3\cr
$^{68}Zn$&14.1& 2.7&-0.3&-12.8\cr
$^{72}Kr$& 8.6&-0.7&-5.0&-20.2\cr
$^{72}Se$&15.0& 5.7& 1.4&-14.5\cr 
$^{74}Kr$& 5.0&-1.4&-4.8&-25.9\cr
\cr
$^{62}Cu$&19.9&1.6& 1.1& 4.9\cr
$^{62}Co$&15.4&2.7& 0.8& 1.5\cr
$^{64}Ga$&23.3&3.8&-0.9& 5.6\cr
$^{64}Cu$&18.8&3.0&-0.8& 1.1\cr
$^{64}Co$&14.1&3.7& 1.2& 2.4\cr
$^{66}Ga$&22.7&5.5&-0.3& 2.2\cr
$^{66}Cu$&15.9&3.7&-0.4&-5.0\cr
$^{68}Ga$&17.8&5.5& 0.0&-6.3\cr
$^{68}Cu$&10.8&3.0&-0.5&-14.1\cr
$^{74}Br$&14.0&7.9& 3.8&-18.3\cr
\cr
\noalign{\hrule}\cr}$$

\newpage

\begin{center}
{\bf Table 2}
\end{center}

Calculated occupancies by SDM for the fp-shell nuclei in the upper half.

$$\halign{&\hfil\  # \ \hfil \ \cr
\noalign{\hrule}\cr
Atomic&Number of&&\multispan{4}\hfil Occupancy  \hfil \cr
Number&valence&Isospin&$f_{7/2}$&$f_{5/2}$&$p_{3/2}$&$p_{1/2}$
\cr
&particles\cr
\noalign{\hrule}\cr
64&24&0&14.59&2.41&5.20&1.81\cr
  &  &4&14.40&3.24&4.58&1.77\cr
  \cr
69&29&5&15.38&4.60&6.44&2.59 \cr
  & &11&15.15&6.19&5.38&2.27 \cr
\cr
74&34&1&15.96&6.62&7.83&3.60 \cr
  &  &2&15.00&3.00&4.00&2.00 \cr
\cr
\noalign{\hrule}\cr}$$

\newpage

\begin{center}
{\bf Table 3}
\end{center}

The sum   rule strength for Gamow - Teller (GT) transition for T = 0
nuclei in $fp$ shell with KB3 interaction by spectral distribution using
Eq.(3) which includes terms up to CLT(column B). Column A gives values with 
only the first term of Eq.(3).

$$\halign{& \hfil\  # \ \hfil \  \cr                                        
\noalign{\hrule}\cr
no. of &\multispan{2}\hfil GT sum rule strength \hfil\cr
valence&\multispan{2}\hrulefill\cr
particles&A&B \cr
\noalign{\hrule}\cr
22 &14.89 &14.46 \cr
\cr
24 &14.44 &14.03 \cr
\cr
32 & 9.62 & 9.44 \cr
\cr
34 & 7.67 & 7.56 \cr
\cr
\noalign{\hrule}\cr}$$

\end{document}